\documentclass[12pt]{article}
\usepackage{graphicx}
\usepackage{lineno}

\newcommand\pubnumber{}
\newcommand\pubdate{\today}

\def\institute{DESY\\
Notkestrasse 85, 22607 Hamburg. Germany}
\def\support{\footnote{Work supported by the Helmholtz Association of German Research Centres.}}

\def\Title#1{\begin{center} {\Large #1 } \end{center}}
\def\Author#1{\begin{center}{ \sc #1} \end{center}}
\def\Address#1{\begin{center}{ \it #1} \end{center}}

\newcommand\pubblock{\rightline{\begin{tabular}{l} \pubnumber\\
         \pubdate  \end{tabular}}}
\newenvironment{Abstract}{\begin{quotation}  }{\end{quotation}}
\newenvironment{Presented}{\begin{quotation} \begin{center} 
             PRESENTED AT\end{center}\bigskip 
      \begin{center}\begin{large}}{\end{large}\end{center} \end{quotation}}

\def\beq{\begin{equation}}
\def\eeq#1{\label{#1}\end{equation}}
\def\eeqn{\end{equation}}

\def\beqa{\begin{eqnarray}}
\def\eeqa#1{\label{#1}\end{eqnarray}}
\def\eeqan{\end{eqnarray}}

\let\bar=\overbar

\def\Dslash{\not{\hbox{\kern-4pt $D$}}}
\def\dslash{\not{\hbox{\kern-2pt $\del$}}}

\def\msb{{\bar{\ssstyle M \kern -1pt S}}}

\begin{document}
\begin{titlepage}
\pubblock

\vfill
\Title{Measurements of the charge asymmetry in top-quark pair production in the dilepton final state at $\sqrt{s}=8$ TeV with the ATLAS detector}
\vfill
\Author{ Roger Naranjo\support~on behalf of the ATLAS Collaboration}
\Address{\institute}
\vfill
\begin{Abstract}
 Measurements of the top--antitop quark pair production charge asymmetry
    in the dilepton channel, characterized by two high-${p}_{\rm{T}}$ leptons (electrons or
muons), are presented using data
    corresponding to an integrated luminosity of  20.3 $\textrm{fb}^{-1}$ from $pp$
    collisions at a center-of-mass energy $\sqrt{s} = 8$~TeV collected
    with the ATLAS detector at the Large Hadron Collider
    at CERN. Inclusive and differential measurements as a function of
    the invariant mass, transverse momentum, and longitudinal boost of
    the $t\bar{t}$  system
    are performed both in the full phase space and in a fiducial
    phase space closely matching the detector acceptance.
    Two observables are studied: $A_C^{\ell \ell}$ based on the selected leptons and $A_C^{t\bar{t}}$ based on the
    reconstructed $t\bar{t}$ final state. No significant deviation from the Standard Model expectations is observed.
\end{Abstract}
\vfill
\begin{Presented}
$9^{th}$ International Workshop on Top Quark Physics\\
Olomouc, Czech Republic,  September 19--23, 2016
\end{Presented}
\vfill
\end{titlepage}
\def\thefootnote{\fnsymbol{footnote}}
\setcounter{footnote}{0}

\section{Introduction}

The measurements of the charge asymmetry provide a good precision test of the Standard Model (SM).  In the SM, the asymmetry is produced by interferences between the Born and one-loop diagram of the $q\bar{q} \rightarrow t\bar{t}$  ̄processes and between
$q\bar{q} \rightarrow t\bar{t}g$ diagrams with initial-state and final-state radiation. In the $t\bar{t}$ rest frame, this asymmetry causes the top quark to be preferentially emitted in the
direction of the initial quark, and causes the antitop quark to be emitted in the direction of the initial
antiquark. In the $pp$ collision at the LHC, valence quarks carry on average a larger fraction of the proton
momentum than sea antiquarks, hence top antiquarks produced through $q\bar{q}$ annihilation are
more central than top quarks. In dileptonic events, the charge asymmetry can be measured in two
 complementary ways: using the pseudorapidity of the charged leptons  or
using the rapidity of the top quarks.
The asymmetry based on the charged leptons uses the difference of the absolute pseudorapidity values of the positively and negatively charged leptons, $|\eta_{\ell^{+}}|$ and
$|\eta_{\ell^{-}}|$. The leptonic asymmetry is defined as
\begin{linenomath}
\begin{equation}
\label{eq:ac_lep}
A_{\rm{C}}^{\ell\ell} = \frac{N(\Delta |\eta| >0) - N(\Delta |\eta| <0)}{N(\Delta |\eta|
  >0) + N(\Delta |\eta| <0)} \, \, \, \textrm{with} \, \, \, \Delta |\eta| = |\eta_{\ell^{+}}|-|\eta_{\ell^{-}}|,
\end{equation}
\end{linenomath}
where $N(\Delta |\eta| >0)$ and $N(\Delta |\eta| <0)$ represent the number of events with
positive and negative $\Delta |\eta|$, respectively.
For the $t\bar{t}$ charge asymmetry the absolute values of the  top and antitop quark rapidities ($|y_{t}|$ and $|y_{\bar{t}}|$, respectively) are used. The $t\bar{t}$ charge asymmetry is defined as
\begin{linenomath}
\begin{equation}
\label{eq:ac}
A_{\rm{C}}^{t\bar{t}}= \frac{N(\Delta |y| > 0) - N(\Delta |y| < 0)}{N(\Delta |y|
  >0) + N(\Delta |y| <0)} \, \, \, \textrm{with} \, \, \, \Delta |y| = |y_{t}|-|y_{\bar{t}}|,
\end{equation}
\end{linenomath}
where $N(\Delta |y| > 0)$ and $N(\Delta |y| < 0)$ represent the number of events with
positive and negative $\Delta |y|$, respectively. 

In these proceedings, the inclusive and differential measurements of the leptonic and $t\bar{t}$ charge asymmetry  in the dilepton channel using  data collected by the ATLAS detector~\cite{atlas} corresponding to an integrated luminosity of  20.3 $\textrm{fb}^{-1}$ from $pp$ collisions at a center-of-mass energy $\sqrt{s} = 8$~TeV are presented~\cite{Aad:2016ove}. The differential measurements are performed as a function of the mass ($m_{t\bar{t}}$), transverse momentum ($p_T^{t\bar{t}}$) and boost ($\beta_{z}^{t\bar{t}}$) of the $t\bar{t}$ system. The measurements are performed in a fiducial region and in the full phase space.
\section{Event Selection and Reconstruction}

Events are required to have exactly two leptons of opposite electric charge and at least two jets with $p_{T}>25$ GeV within $|\eta|<2.5$. In all three final states, exactly two isolated leptons with opposite
charge and an invariant mass $m_{\ell  \ell} > 15$ GeV are required.
In the same-flavor channels ($ee$ and $\mu\mu$),
the invariant mass of the two charged leptons is required to be
outside of the $Z$ boson mass window such
that \mbox{$|m_{\ell \ell}-m_Z|>10$  GeV}.
Furthermore, it is required that missing transverse momentum is greater then $30$ GeV and at
least one of the jets must be $b$-tagged. In the $e\mu$ channel,  the scalar sum of the $p_{T}$ of the two leading jets and
leptons is required to be larger than $130$ GeV.

The main background contribution comes from Drell--Yan production
of $Z/\gamma^{*} \rightarrow \ell \ell$, which is estimated by a combination of simulated samples and corrections
derived from data. The smaller contributions from diboson
and single-top-quark production are evaluated purely via simulations. Contributions arising from  events including a jet or a lepton from a semileptonic hadron decay
misidentified as an isolated charged lepton as well as leptons from photon conversions, are estimated using simulated samples, modified with corrections
derived from data.

\begin{figure}[htb]
\centering
\includegraphics[height=1.8in]{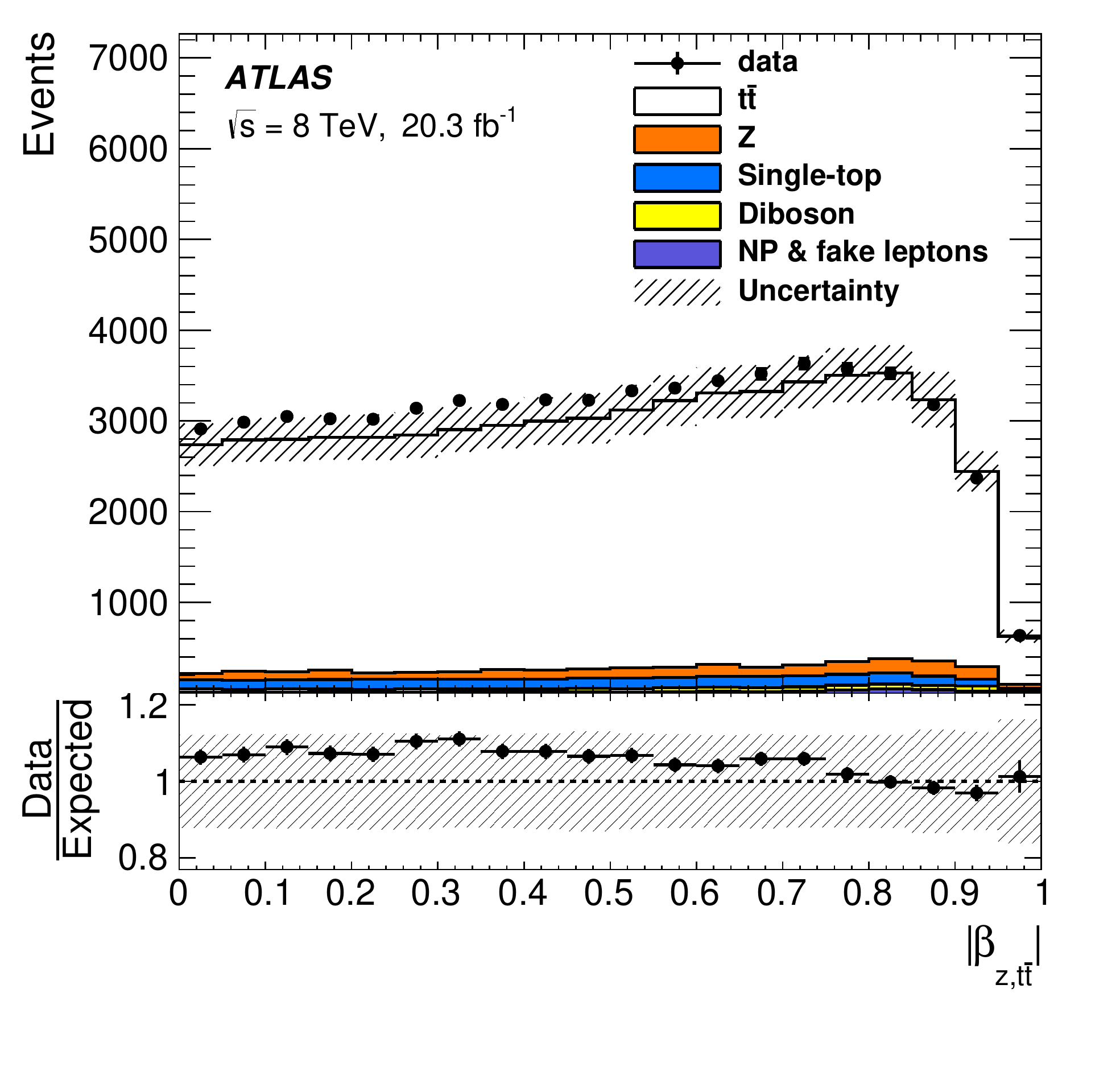}
\includegraphics[height=1.8in]{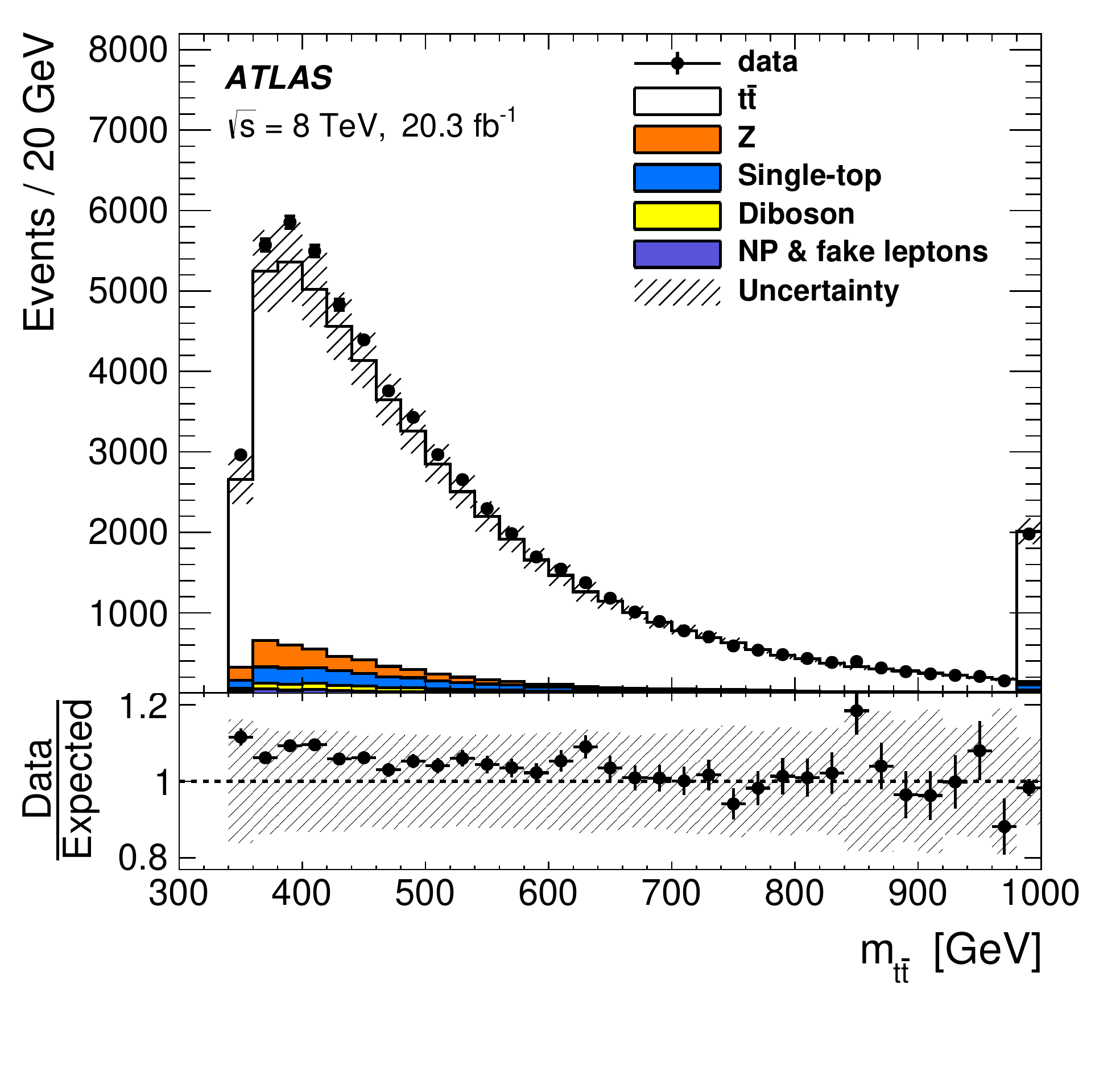}
\includegraphics[height=1.8in]{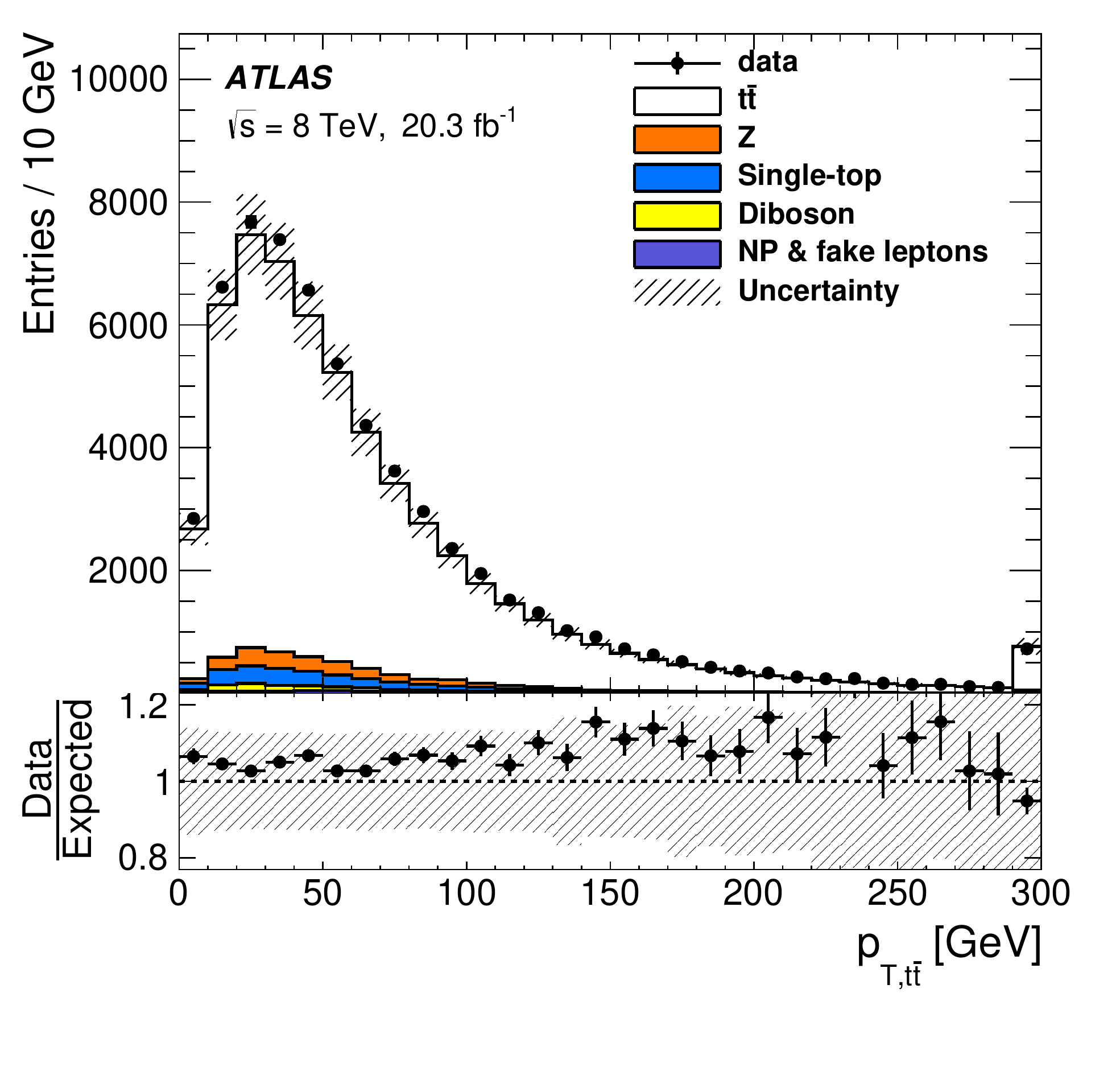}
\caption{Comparison between observations and expectations for the boost (left) mass (middle), and $p_T$ of the $t\bar{t}$ system. The total uncertainty on the distributions are shown~\cite{Aad:2016ove}. }
\label{fig:magnet}
\end{figure}

The $t\bar{t}$ system is reconstructed in order to perform the inclusive and differential measurements of $A_C^{t\bar{t}}$. The system is reconstructed using the KIN method. The KIN method assumes the mass of the top quarks (172.5 GeV) and $W$ mass (80.4 GeV), and solve the system of equations obtained from momentum convervation numerically using the Newton-Rhapson method. The reconstruction efficiency is above 90\%. A comparison between observations and expectations is shown in Fig.~\ref{fig:magnet} after event reconstruction. A good agreement within the uncertainties is observed.

\section{Unfolding}
The measurements of $A_C^{\ell \ell}$ and  $A_C^{t\bar{t}}$ are corrected in order to remove the effects introduced by the detector. This correction is performed by using the Fully Bayesian Unfolding (FBU) method~\cite{fbu}. The measurements are unfolded back to a stable particle level in a fiducial region closely matching the detector acceptance, and back to parton level in the full phase space. The combined measurement of the three decay channels is performed during the unfolding procedure. The asymmetries are computed using the posterior probability density obtained as an output of the unfolding procedure. Systematic uncertainties related with detector modeling and background modeling are evaluated during the unfolding by using a marginalization procedure.

\section{Results}
\begin{figure}[htb]
\centering
\includegraphics[height=2.0in]{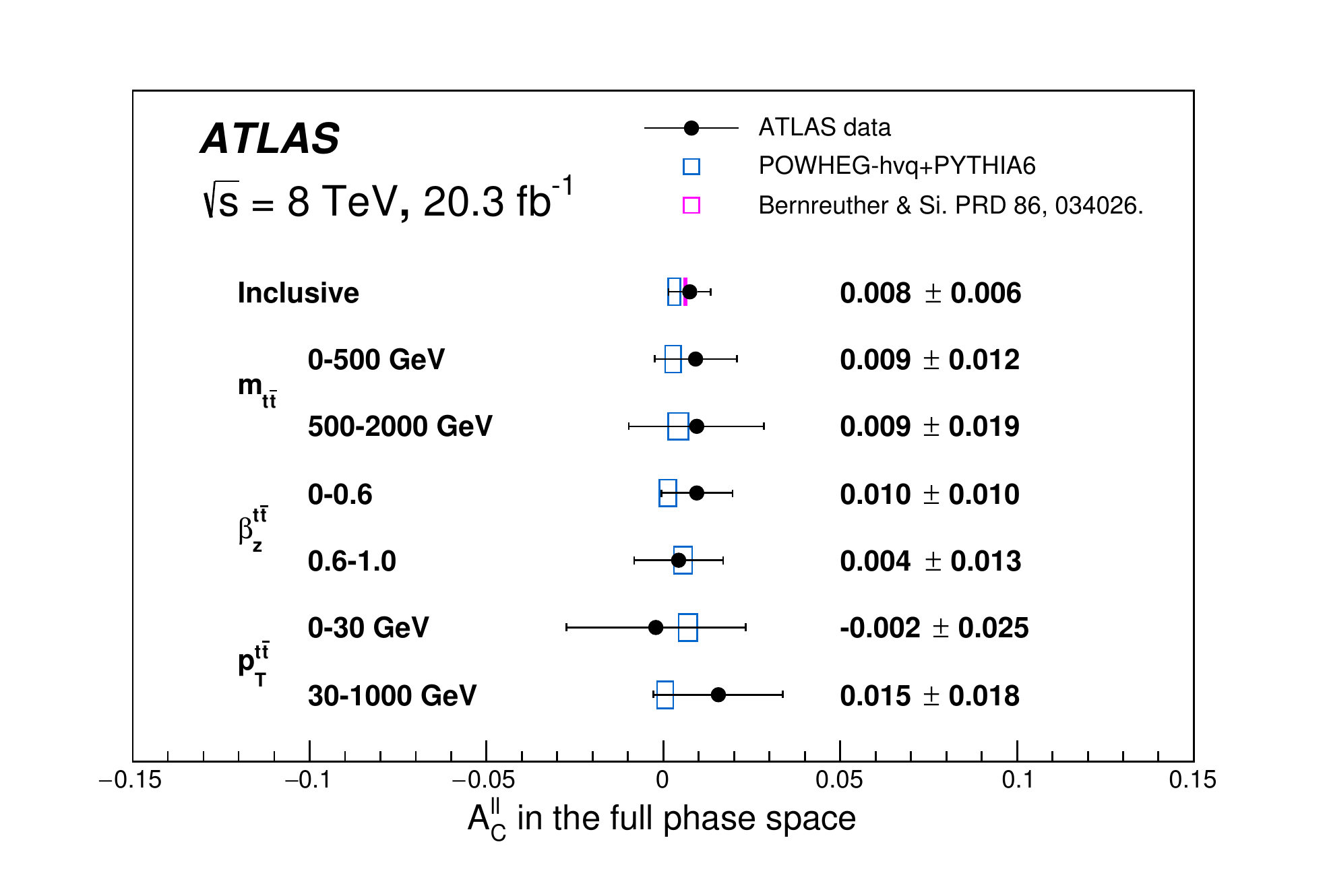}
\includegraphics[height=2.0in]{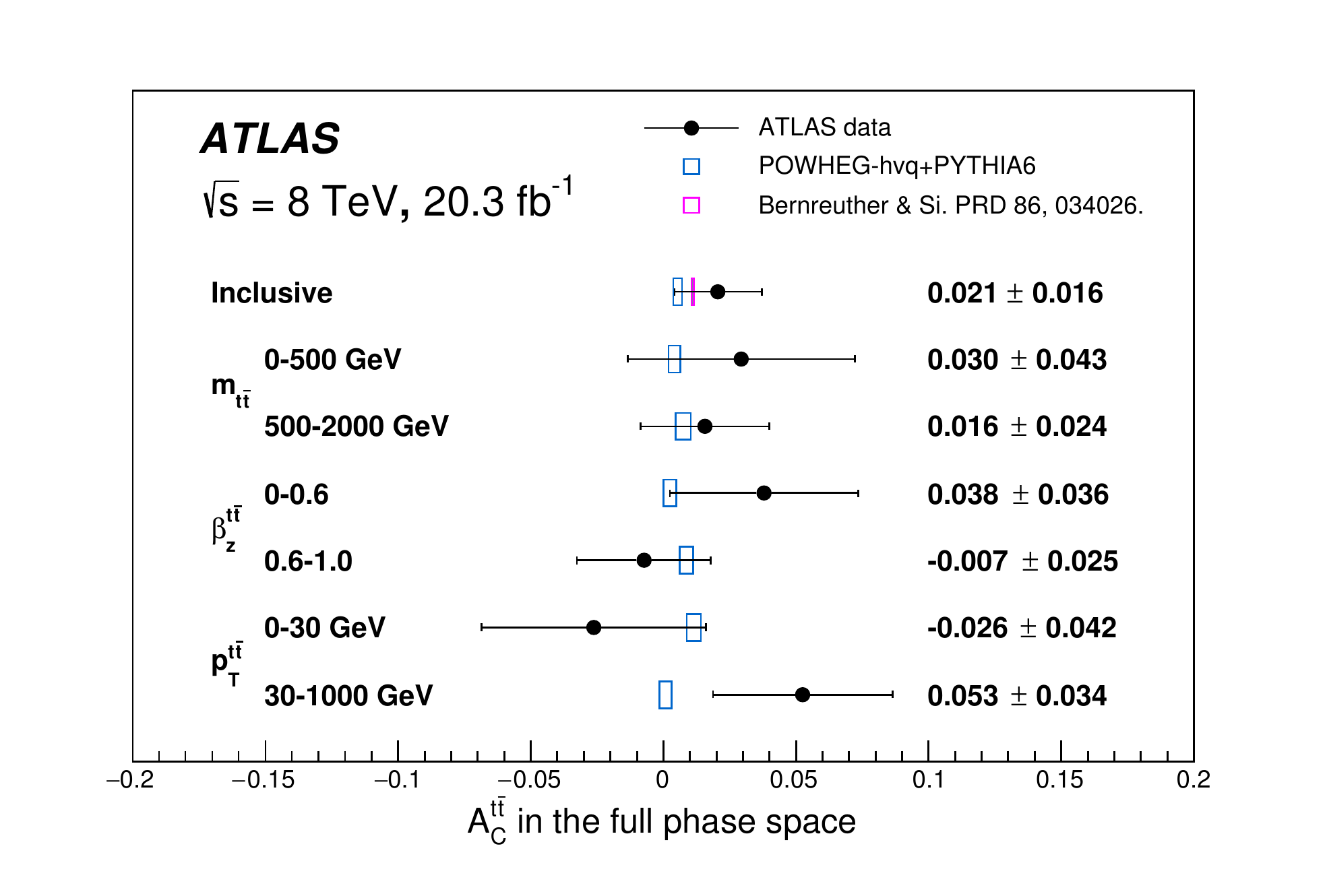}

\caption{Summary of the inclusive and differential measurements of the $t\bar{t}$ asymmetry (left) and lepton asymmetry (right) performed in the full phase space. The measurements are compared with theoretical predictions~\cite{Aad:2016ove}.}
\label{fig:sumary}
\end{figure}

\begin{figure}[htb]
\centering
\includegraphics[height=1.8in]{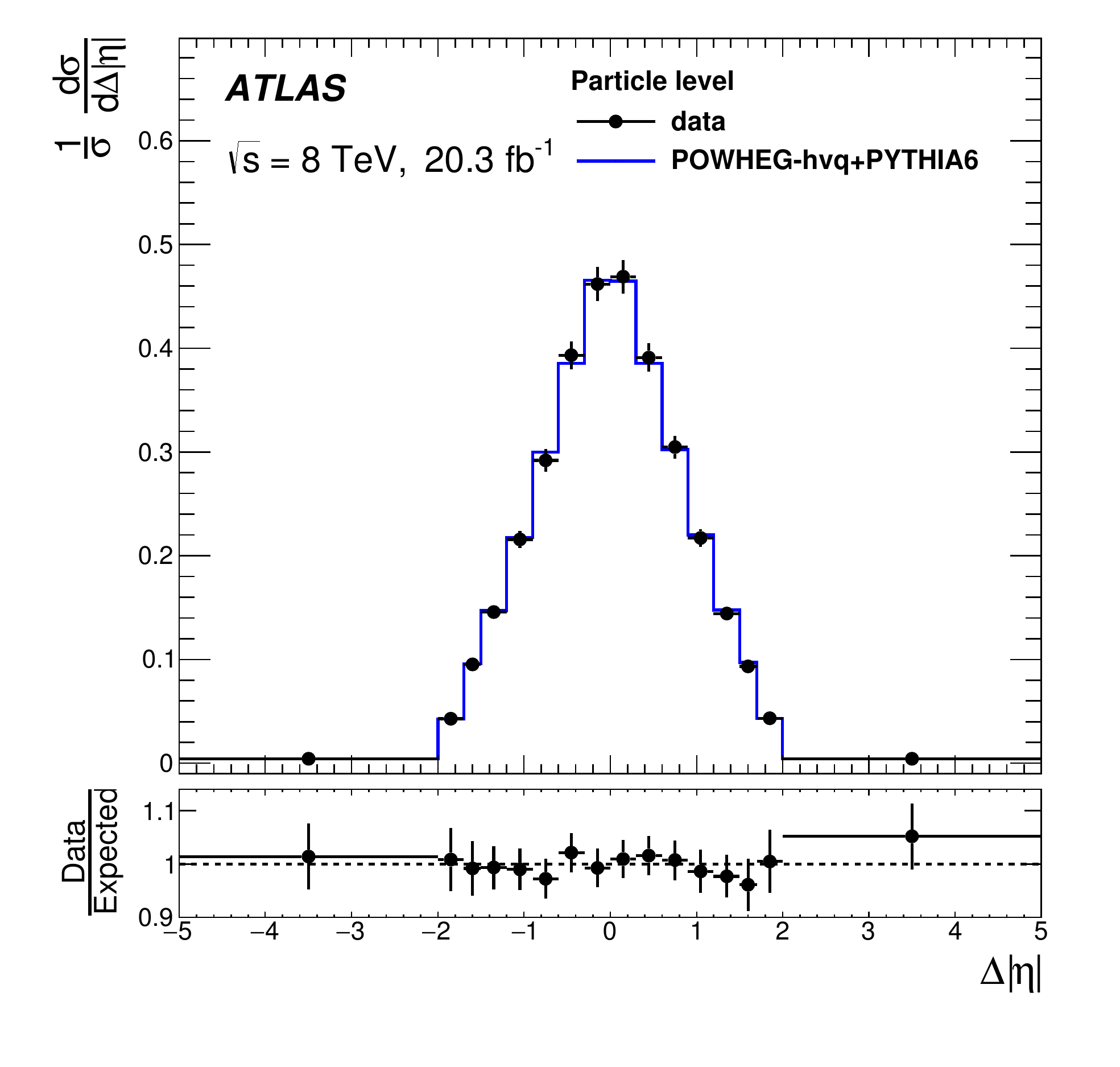}
\includegraphics[height=1.8in]{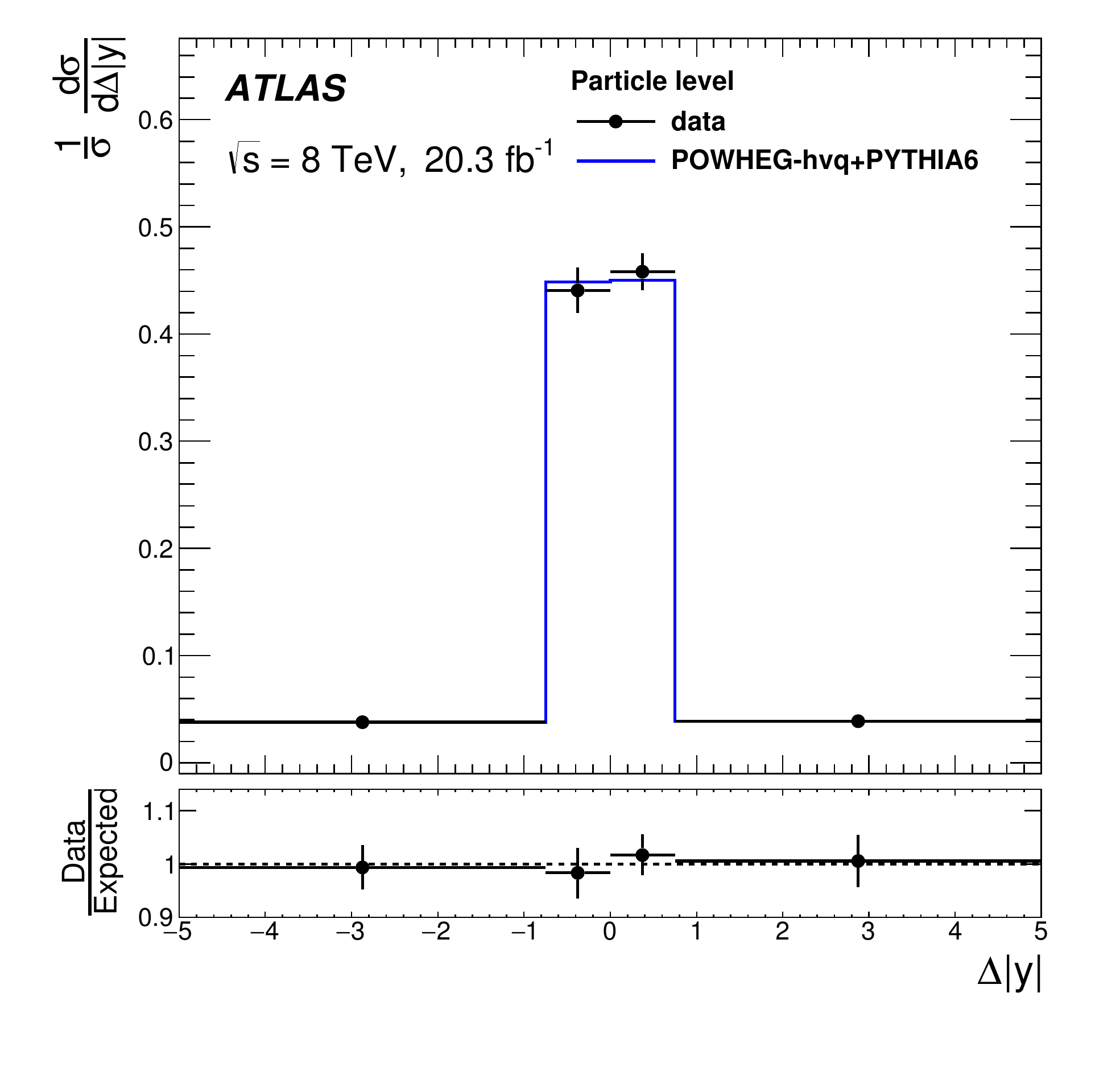}

\caption{Data distribution after the unfolding procedure compared with the prediction for the inclusive $\Delta|\eta|$ (left) and $\Delta|y|$ (right) observables in the fiducial volume. The data/expected ratio is also
shown~\cite{Aad:2016ove}.}
\label{fig:distribution}
\end{figure}

\begin{figure}[htb]
\centering
\includegraphics[height=2.0in]{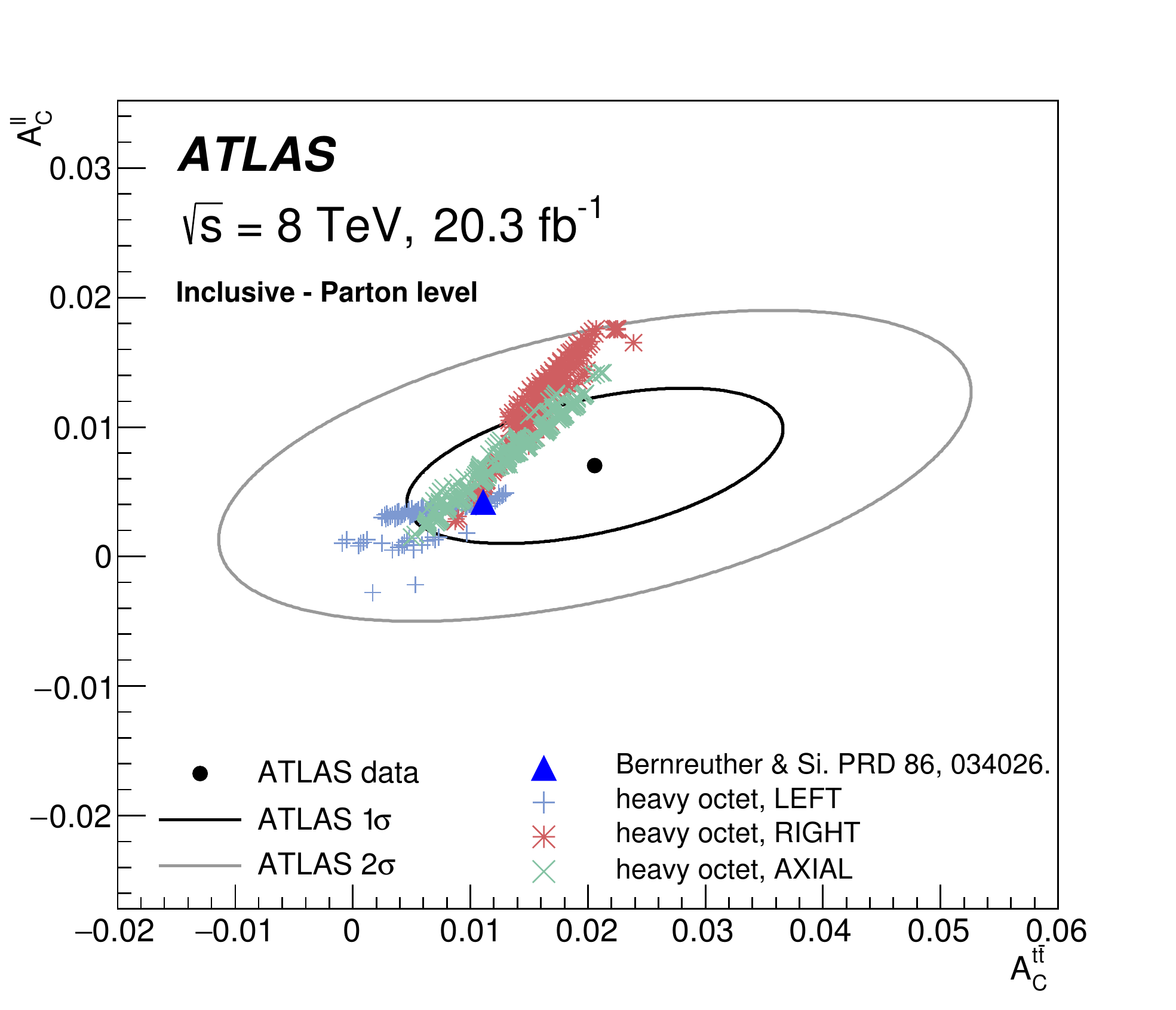}
\includegraphics[height=2.0in]{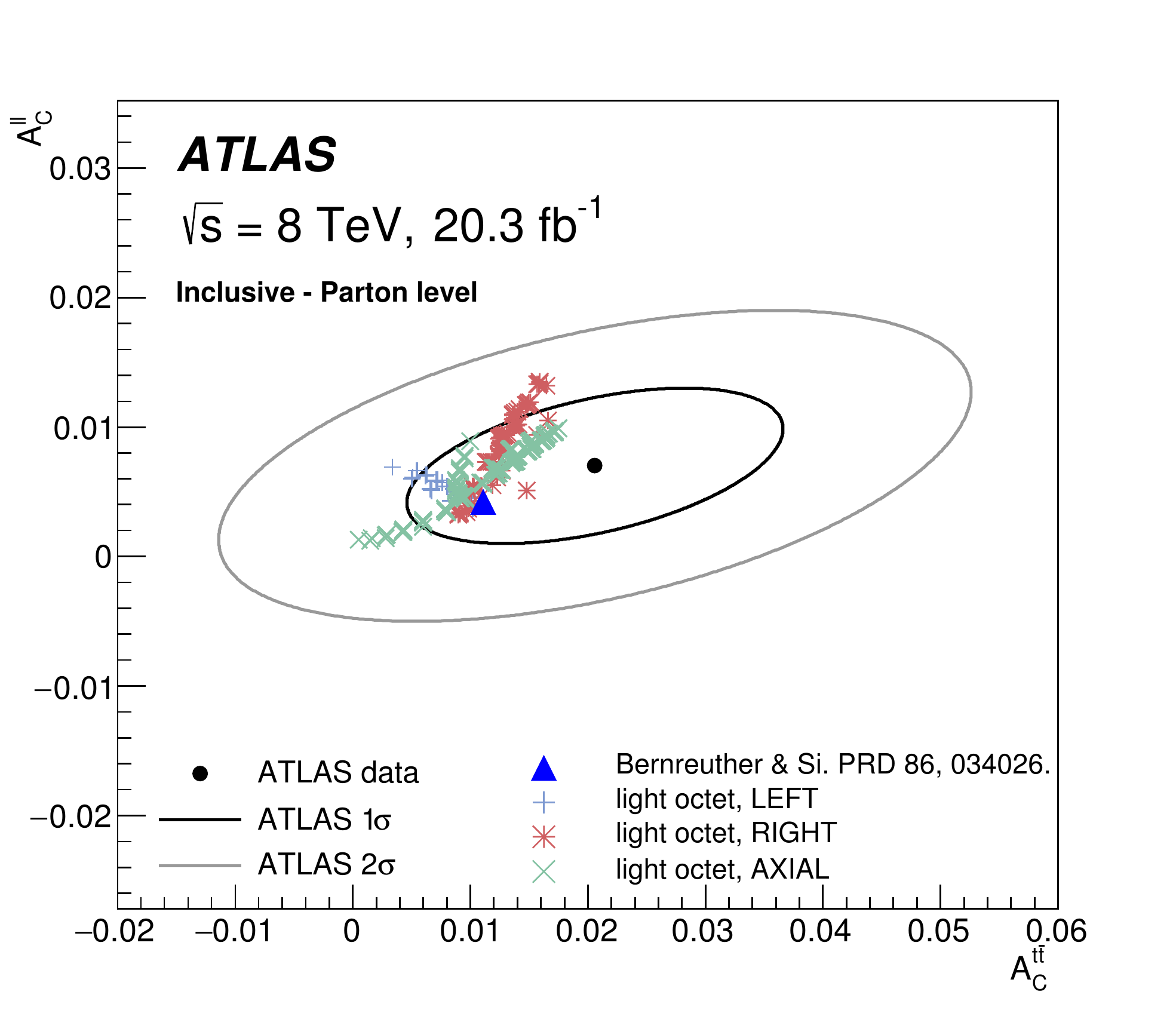}

\caption{Comparison of the inclusive $A_C^{\ell \ell}$ and $A_C^{t\bar{t}}$ measurement values in the full phase space to the SM  and to two benchmark BSM models~\cite{Aad:2016ove}. }
\label{fig:2d}
\end{figure}

Figure~\ref{fig:sumary} shows the inclusive and differential measurements performed for $A_C^{\ell \ell}$ and  $A_C^{t\bar{t}}$ at parton level in the full phase space. The total uncertainty on the measurements is shown. The main source of uncertainty on the different measurements is the statistical uncertainty, followed by the signal modeling uncertainty. The measurements that involve the reconstruction of the $t\bar{t}$ system are also affected by a reconstruction uncertainty which is approximately half of the size of the statistical uncertainty. The uncertainties corresponding to the detector and background modeling do not contribute significantly to the total uncertainty. The results are compatible with the SM predictions~\cite{Bernreuther:2012sx}. A similar behavior is observed on the uncertainties in the measurements performed in the fiducial region, however, there is a reduction in the modeling uncertainties. Figure~\ref{fig:distribution} shows the unfolded distribution for the $\Delta|y|$ and $\Delta|\eta|$ observables in the fiducial region. The distribution is in agreement with SM predictions. Figure~\ref{fig:2d} shows the $A_C^{\ell \ell}$ and $A_C^{t\bar{t}}$ measurements in comparison with several models beyond the SM~\cite{Aguilar-Saavedra:2014nja} in the full phase space. In these models, the values of the asymmetry are expected to be different from the SM expectations. The ellipses correspond to the $1\sigma$ and $2\sigma$ total uncertainty on the measurements. The correlation between $A_C^{\ell \ell}$ and $A_C^{t\bar{t}}$ is about 48\%. The measurements are compatible with the SM and do not exclude the two sets of BSM models considered.

\end{document}